\documentclass[aps,prl,twocolumn,superscriptaddress,a4paper,showpacs]{revtex4}
\usepackage{amsmath}
\usepackage{graphicx}
\begin{document}
\title{Duality Relation for Quantum Ratchets}
\author{J. Peguiron}
\affiliation{Kavli Institute of Nanoscience, Delft University of Technology, Lorentzweg 1, 2628 CJ Delft, The
Netherlands} \affiliation{Institut f\"ur Theoretische Physik, Universit\"at Regensburg, D-93040 Regensburg,
Germany}
\author{M. Grifoni}
\affiliation{Institut f\"ur Theoretische Physik, Universit\"at
Regensburg, D-93040 Regensburg, Germany}
\date{\today}
\begin{abstract}
A duality relation between the long-time dynamics of a quantum Brownian particle in a tilted ratchet potential
and a driven dissipative tight-binding model is reported. It relates a situation of weak dissipation in one
model to strong dissipation in the other one, and vice versa. We apply this duality relation to investigate
transport and rectification in ratchet potentials: From the linear mobility we infer ground-state
\textit{delocalization} for weak dissipation. We report reversals induced by adiabatic driving and temperature
in the ratchet current and its dependence on the potential shape.
\end{abstract}
\pacs{05.30.-d, 05.40.-a, 73.23.-b, 05.60.Gg}
\maketitle
%
%     Introduction
%
Periodic structures with broken spatial symmetry, known as ratchet systems~\cite{RatRev}, present the attractive
property of allowing transport under the influence of unbiased forces. The interplay of dissipative
tunneling~\cite{WeiBK99} with unbiased driving enriches the quantum ratchet effect with features absent in its
classical counterpart like, e.g., current reversals as a function of temperature~\cite{ReiPRL97,LinSci99}.
Quantum ratchet systems have only recently been experimentally realized in semiconductor~\cite{LinSci99} and
superconductor~\cite{MajPRL03} devices. Also from the theory side there are still few
works~\cite{ReiPRL97,SchPRB02,LehPRL02,RonPRL98,GriPRL02,MacPRE04} which, with the exception
of~\cite{SchPRB02,LehPRL02}, are restricted to the regime of moderate-to-strong damping. After the pioneering
semiclassical work~\cite{ReiPRL97}, further progress towards a quantum description was made in~\cite{GriPRL02},
where the role of the band structure in ratchet potentials sustaining few bands below the barrier was
investigated. Recently, a quantum Smoluchowski treatment~\cite{MacPRE04} added to the available methods. In this
paper, we generalize to an arbitrary ratchet potential a duality relation put forward in~\cite{FisPRB85} for a
cosine potential. It provides a tight-binding description of quantum Brownian motion in a ratchet potential, and
leads to an expression for the ratchet current valid in a wide parameter range including weak dissipation and
nonlinear adiabatic driving. We apply this method to discuss rectification and ground-state delocalization
occurring for \textit{weak dissipation} in ratchet potentials. Our results encompass correctly the classical
limit.

%
%     Duality transformation
%
We consider the Hamiltonian $\hat{H}_\text{R}$ of a quantum particle
of mass $M$
in a one-dimensional periodic potential $V{(q+L)}=V(q)$ tilted by a
time-dependent force $F(t)$,
\begin{equation}
\hat{H}_{\text{R}}(t)=\frac{\hat{p}^2}{2M}+V(\hat{q})-F(t)\hat{q}.
\end{equation}
The potential assumes in Fourier expansion the form
\begin{equation}\label{EqnPot}
V(\hat{q})=\sum\nolimits_{l=1}^{\infty}{V_{l}\cos{\left(2\pi l\hat{q}/L-\varphi_{l}\right)}},
\end{equation}
and can take any shape. Apart from special configurations $\lbrace V_{l}\sin{(\varphi_{l}-l\varphi_{1})}=0\
\forall l\rbrace$ of the amplitudes~$V_{l}$ and phases~$\varphi_{l}$, this potential is spatially asymmetric and
describes a ratchet system. The interaction of the system with a dissipative thermal environment is modeled by
the standard Hamiltonian~$\hat{H}_{\text{B}}$ of a bath of harmonic oscillators whose coordinates are bilinearly
coupled to the system coordinate~$\hat{q}$~\cite{WeiBK99}. The bath is fully characterized by its spectral
density $J(\omega)$. We consider strict Ohmic damping $J(\omega)=\eta\omega$, which reduces to instantaneous
viscous damping (viscosity~$\eta$) in the classical limit. In such a system, the ratchet effect is characterized
by a nonvanishing average stationary particle current $v_\text{R}^{\infty}
=\lim_{t\to\infty}t^{-1}\int_{0}^{t}dt^{\prime}v(t^{\prime})$ in the presence of unbiased driving, characterized
by $\lim_{t\to\infty}t^{-1}\int_{0}^{t}dt^{\prime}F(t^{\prime})=0$, switched on at time $t^{\prime}=0$. In this
paper, we shall consider the particular case of unbiased bistable driving switching adiabatically between the
values~$\pm F$. We report a method to evaluate the stationary velocity~$v^{\infty}_\text{DC}(F)$ in the biased
situation of time-independent driving $F$, which is also of experimental interest~\cite{MajPRL03,LinSci99}. The
ratchet current in the presence of adiabatic bistable driving can be expressed as $v_\text{R}^\infty=
v^{\infty}_\text{DC}(F)+v^{\infty}_\text{DC}(-F)$.
\begin{figure}
\includegraphics{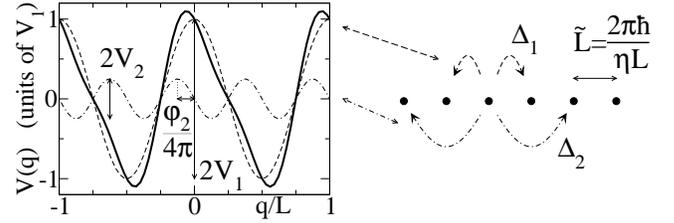} \caption{\label{Fig1}Dual relation between a dissipative ratchet system and a
tight-binding (TB) model sketched for a two-harmonics ratchet potential (thick curve). Each harmonic (thin
curves) generates couplings to different neighbors in the TB system, according to Eqs.~(\ref{EqnRelDelta}) and
(\ref{EqnHTB}). The periodicity~$\tilde{L}$ of the TB model is determined by the viscosity $\eta$ in the
original model.}
\end{figure}

The whole information on the system dynamics is contained in the reduced density matrix
$\hat{\rho}(t)=\text{Tr}_{\text{B}}\hat{W}(t)$, obtained from the density matrix $\hat{W}(t)$ of the
system-plus-bath $\hat{H}=\hat{H}_{\text{R}}+\hat{H}_{\text{B}}$, with time-independent driving~$F$, by
performing the trace over the bath degrees of freedom. To evaluate the evolution of the average position
$\langle q(t)\rangle=\text{Tr}_{\text{R}}\lbrace\hat{q}\hat{\rho}(t)\rbrace$, the diagonal
elements~$P(q,t)=\langle q|\hat{\rho}(t)|q\rangle$ of the reduced density matrix are needed, and can be obtained
by real-time path integrals techniques~\cite{WeiBK99}. The velocity follows by time differentiation. At initial
time~$t^{\prime}=0$, we assume a preparation in a product form
 where the bath is in thermal equilibrium with the ratchet system
$\hat{W}(0)=\hat{\rho}(0)e^{-\beta\hat{H}_{\text{B}}}
\lbrack\text{Tr}_{\text{B}}e^{-\beta\hat{H}_{\text{B}}}\rbrack^{-1}$. The bath temperature is fixed by
$T=1/\beta k_{\text{B}}$. This leads to a double path integral
\begin{eqnarray}
\lefteqn{P(q_{\text{f}},t)=}\\
&&\int dq_{\text{i}}\int dq_{\text{i}^{\prime}}
\langle q_{\text{i}}|\hat{\rho}(0)|q_{\text{i}^{\prime}}\rangle
\int\limits_{q_{\text{i}}}^{q_{\text{f}}}\mathcal{D}q
\int\limits_{q_{\text{i}}^{\prime}}^{q_{\text{f}}}\mathcal{D}^*q^{\prime}
A\lbrack q\rbrack A^*\lbrack q^{\prime}\rbrack
F\lbrack q,q^{\prime}\rbrack\nonumber
\end{eqnarray}
on the continuous coordinates $q$ and $q^{\prime}$. Here $A\lbrack
q\rbrack=\exp{\lbrace-(it/\hbar)\hat{H}_{\text{R}}\rbrace}$ is the propagator of the ratchet system for a
path~$q(t^{\prime})$, and $F\lbrack q,q^{\prime}\rbrack$ the Feynman-Vernon influence functional of the bath
inducing nonlocal-in-time Gaussian correlations between the paths $q(t^{\prime})$ and
$q^\prime(t^{\prime})$~\cite{WeiBK99}. Due to the nonlinearity of the potential~$V(q)$, these path integrals
cannot be performed explicitly. For a cosine potential, \citet*{FisPRB85} introduced an exact expansion in the
propagator~$A\lbrack q\rbrack$ which transforms the path integrals into Gaussian ones that can be performed.
Generalizing this idea for the arbitrary periodic potential~(\ref{EqnPot}), we find the expansion
\begin{widetext}
\begin{equation}\label{EqnExpPot}
\exp\left\lbrace-\frac{i}{\hbar}\int\limits_{0}^{t}dt^{\prime} V\left(q(t^{\prime})\right)\right\rbrace=
\sum_{m=0}^{\infty}\sum_{\lbrace l_{j}\rbrace} \prod_{j=1}^{m}\left(\frac{-i\Delta_{l_{j}}}{\hbar}\right)
\int\limits_{0}^{t}dt_{m}\int\limits_{0}^{t_{m}}dt_{m-1} \ldots\int\limits_{0}^{t_{2}}dt_{1}
\exp\left\lbrace-\frac{i}{\hbar}\int\limits_{0}^{t}dt^{\prime} \rho(t^{\prime})q(t^{\prime})\right\rbrace,
\end{equation}
\end{widetext}
where
$\rho(t^{\prime})=(2\pi\hbar/L)\sum_{j=1}^{m}l_{j}\delta(t^{\prime}-t_{j})$,
and
\begin{equation}\label{EqnRelDelta}
\Delta_{l}=\frac{V_{l}}{2}e^{i\varphi_{l}} \quad\text{for }l>0,\quad\Delta_{-l}=\Delta_{l}^{*}.
\end{equation}
The physical meaning of these new quantities will be discussed
later. For each term of the sum on~$m$ in~(\ref{EqnExpPot}) we
have introduced $m$~intermediate
 times~$t_{j}$,
and corresponding indices~$l_{j}$ taking any value among
 $\lbrace\pm 1,\pm 2,\ldots\rbrace$. The sum~$\sum_{\lbrace l_{j}\rbrace}$
 runs on all configurations of these indices. A similar expansion
is performed for the propagator~$A^*\lbrack q^{\prime}\rbrack$,
involving a new set of $m^{\prime}$~times~$t_j^{\prime}$ and
 indices~$l_j^{\prime}$ being used to define
$\rho^{\prime}(t^{\prime})$ similarly to~$\rho(t^{\prime})$. This
enables us to rewrite the average position $\langle
q(t)\rangle=\int dqqP(q,t)$ in terms of a series in the
amplitudes~$V_{l}$ of the potential.

Though still intricate, the resulting expression becomes easier to treat in the long-time limit we are
interested in. Quantitatively, the measurement time~$t$ should be very long on the time scale
$\gamma^{-1}=(\eta/M)^{-1}$ set by dissipation. A second approximation is necessary to proceed: we neglect terms
$e^{-\gamma t_{j}}$, $e^{-\gamma t_{j}^{\prime}}$, $e^{-\gamma(t-t_{j})}$, $e^{-\gamma(t-t_{j}^{\prime})}$,
$e^{-\omega_{\text{B}}t_{j}}$, $e^{-\omega_{\text{B}}t_{j}^{\prime}}$, $e^{-\omega_{\text{B}}(t-t_{j})}$, and
$e^{-\omega_{\text{B}}(t-t_{j}^{\prime})}$, where $\omega_{\text{B}}=2\pi k_{\text{B}}T/\hbar$, in the
integrands involved in the series expression for~$\langle q(t)\rangle$. We shall refer to this assumption as the
rare transitions (RT) limit and discuss its validity later. Generalizing~\cite*{FisPRB85}, we consider a
Gaussian wave packet centered at position~$q_0=\text{Tr}_{\text{R}}\lbrack\hat{q}\hat{\rho}(0)\rbrack$ and
momentum $p_0=\text{Tr}_{\text{R}}\lbrack\hat{p}\hat{\rho}(0)\rbrack$ as initial preparation for the ratchet
system. We obtain the important result
%\begin{widetext}
\begin{equation}\label{EqnDualQ}
\langle q(t)\rangle\underset{\genfrac{}{}{0pt}{}{t\to\infty}{\text{RT}}}{\sim}q_0+ \frac{p_0}{\eta}+\frac{F
t}{\eta}-\langle q_{\text{TB}}(t)\rangle_{\text{TB}}.
\end{equation}
%\end{widetext}
Parts of the series expression for~$\langle q(t)\rangle$ has been summed, yielding the first three terms. The
rest can be identified with the series expression for the expectation value of the position operator
$\hat{q}_{\text{TB}}=\tilde{L}\sum_{n=-\infty}^{\infty}n|n\rangle\langle n|$ of a driven tight-binding~(TB)
system, described by the Hamiltonian
\begin{equation}\label{EqnHTB}
\hat{H}_{\text{TB}}=\sum_{n,l=-\infty}^{\infty} (\Delta_{l}|n+l\rangle\langle n| +\Delta_{l}^{*}|n\rangle\langle
n+l|)-F\hat{q}_{\text{TB}},
\end{equation}
and bilinearly coupled to a \textit{different} bath of harmonic oscillators. The spectral density of this bath
$J_{\text{TB}}(\omega)=J(\omega)/[1+\left(\omega/\gamma\right)^2]$ is still Ohmic but presents a cutoff at the
frequency $\gamma$ set by dissipation. At initial time~$t^{\prime}=0$ the TB system is prepared in the
state~$|n=0\rangle$. The calculation shows that the~$\Delta_l$ introduced in~(\ref{EqnRelDelta}) are identified
with the couplings of the TB system~(\ref{EqnHTB}). We stress that the $l\text{th}$ harmonic of the original
potential results in a coupling to the $l\text{th}$ neighbors in the dual TB system as sketched in
Fig.~\ref{Fig1}. One can easily show that the spatial symmetry condition on the phases~$\varphi_{l}$ is the same
in both systems. The first three terms on the right-hand side of~(\ref{EqnDualQ}) reproduce exactly the
classical solution for the average position $\langle q(t)\rangle$ of a free system, $V(q)\equiv 0$, at long
times. In this linear case, the quantum and classical solutions should be identical, due to Ehrenfest theorem,
and they are, because the TB average $\langle q_{\text{TB}}(t)\rangle_{\text{TB}}$ vanishes in the absence of
the potential~$V(q)$. We expect the same result when the potential is present but unimportant, e.g., for large
driving~$F$ and/or high temperatures~$T$.

The series expression for the diagonal elements of the reduced density matrix~$\hat{\rho}_\text{TB}$ of the TB
system, which leads to the series expression for $\langle q_{\text{TB}}(t)\rangle_{\text{TB}}$, can be written
in terms of pairs of TB trajectories $q_\text{TB}(t^{\prime})=\eta^{-1}
\int_{0}^{t^{\prime}}dt^{\prime\prime}\rho(t^{\prime\prime})$ [with $\rho(t^\prime)$ introduced above
Eq.~(\ref{EqnRelDelta})], and $q^\prime_\text{TB}(t^{\prime})$ defined similarly in terms
of~$\rho^\prime(t^\prime)$. From that one extracts the spatial periodicity~$\tilde{L}$ of the TB system,
yielding $\tilde{L}=L/\alpha$, where $\alpha=\eta L^2/2\pi\hbar$ is the dimensionless dissipation parameter of
the original system. These pairs of trajectories combine in discrete paths
in the $q-q^\prime$~plane parametrized
by pairs of integers~$(n,n^{\prime})$.
\begin{figure}
\includegraphics[angle=270]{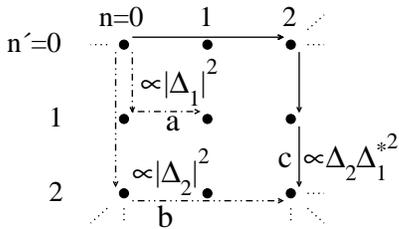} \caption{\label{Fig2}Representation of some of the second-order~(a,b) and
third-order~(c) paths contributing to the diagonal elements of the reduced density matrix of the tight-binding
model, and the corresponding dependence on the couplings~$\Delta_l$.}
\end{figure}
%
%
%.
Each path starting in the diagonal element $(0,0)$ and ending at time~$t$ in $(m,m)$ contributes to $\langle
m|\hat{\rho}_\text{TB}(t)|m\rangle$. Each transition in the path brings a corresponding factor~$\Delta_l$ and
all paths involve at least two transitions~(cf. Fig.~\ref{Fig2}). Written in this form, the diagonal elements of
the reduced density matrix are a solution of a generalized master equation~\cite{GriPRE96} in terms of
transition rates~$\Gamma_{m}$ from the TB site $(n,n)$ to the site $(n+m,n+m)$. Consequently, these rates are
expressed in power series of all the couplings~$\Delta_{l}$, starting from second order. As the times
$t_j,t^\prime_j$ introduced in~(\ref{EqnExpPot}) are identified with the transition times in the TB
representation, the rates~$\Gamma_{m}$ give also a way to control our assumption of rare transitions. It
corresponds to neglect those paths which involve transitions on a time
scale~$\max(\gamma^{-1},\omega_{\text{B}}^{-1})$ after the initial time~$t^{\prime}=0$ or before the final
time~$t^{\prime}=t$. As transitions in the TB model happen on a time scale~$\Gamma_{m}^{-1}$, the duality
relation will be valid when the transitions are rare on the time
scale~$\max(\gamma^{-1},\omega_{\text{B}}^{-1})$, i.e., when all rates satisfy
$\Gamma_m\ll\min(\gamma,\omega_{\text{B}})$. This condition is controlled by the dissipation
through~$\gamma=\eta/M$ and the temperature through~$\omega_{\text{B}}=2\pi k_{\text{B}}T/\hbar$.

%
% Discussion
%
Due to the change of periodicity length between the two systems, the dissipation parameter~$\alpha$ and the
energy drop per unit cell $\epsilon=FL$ become $\tilde{\alpha}=1/\alpha$ and
$\tilde{\epsilon}=\epsilon/{\alpha}$ in the TB system. Thus, weak dissipation in one system maps to strong
dissipation in the other one although the viscosity~$\eta$ in the spectral density does not change. The
asymptotic dynamics is usually described by the nonlinear mobility $\mu=\lim_{t\to\infty}v(t)/F$. With these
notations, the duality relation~(\ref{EqnDualQ}) can be rewritten in the form
\begin{equation}\label{EqnDualMu}
\mu(\alpha,\epsilon)
\xrightarrow[\text{RT}]{}\mu_{0}-\mu_{\text{TB}}(1/\alpha,\epsilon/\alpha),
\end{equation}
where $\mu_{0}=1/\eta$ is the mobility of the free system, $V(q)\equiv 0$. In the special case of a cosine
potential, this relation was already obtained in~\cite{FisPRB85} for the dc mobility. It it also interesting to
notice that it was also derived in~\cite{SasPRB96} for the linear ac mobility in a cosine potential. However, we
did not completely succeed in generalizing Eq.~(\ref{EqnDualMu}) in the presence of time-dependent driving.

\begin{figure}
\includegraphics[angle=270]{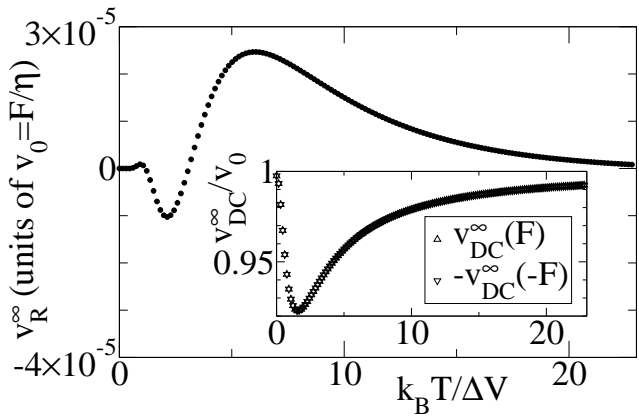} \caption{\label{Fig3}Ratchet current and stationary velocity (inset) as a function of
temperature for the potential of amplitude $\Delta V$ depicted in Fig.~\ref{Fig1}. Weak dissipation is chosen
with $\alpha=0.2$ and $\hbar\gamma=0.76\Delta V$. Driving is set to $FL=0.57\Delta V$.}
\end{figure}

%
% Evaluation of the tight-binding transition rates to 3rd order
%
We shall now focus on the evaluation of the stationary velocity~$v^\infty_\text{DC}(F)$. By solving the
generalized master equation mentioned above, one finds the stationary velocity
$v^\infty_{\text{TB}}=\tilde{L}\sum_{m}m\Gamma_{m}$ in the dissipative TB system. The duality
relation~(\ref{EqnDualQ}) can then be used to obtain
\begin{equation}\label{EqnStatVel}
v^\infty_\text{DC}(F)=F/\eta-(L/\alpha)
\sum\nolimits_{m=1}^{\infty}m(\Gamma_{m}-\Gamma_{-m}).
\end{equation}
As discussed above, the rates~$\Gamma_m$ are power series
in the couplings~$\Delta_l$ starting from second order.
For a given~$m\neq 0$, there are only two possible second-order contributions
to~$\Gamma_m$, which, after use of Eq.~(\ref{EqnRelDelta}), sum up to~\cite{WeiBK99}
\begin{equation}\label{EqnGamma2nd}
\Gamma_{m}^{(2)}=\frac{V_{m}^2}{4\hbar^2\gamma}\int\nolimits_{-\infty}^{\infty}d\tau \
e^{-m^2\tilde{\alpha}Q(\tau)+im(F\tilde{L}/\hbar\gamma)\tau}.
\end{equation}
The influence of the dissipative environment enters through the dimensionless bath correlation function
$Q(\tau)=2\int_{0}^{\infty}\frac{d\omega}{\omega(1+\omega^2)}\left[\coth{(\omega/2\theta)}(1-\cos{\omega\tau)}+i\sin{\omega\tau}\right]$
with $\theta=k_{\text{B}}T/\hbar\gamma$. At zero bias~$F=0$ and in the scaling limit $\hbar\gamma\gg
k_\text{B}T$, the rates show a power-law dependence on temperature $\Gamma_{m}^{(2)}\propto
T^{2m^2\tilde{\alpha}-1}$. The linear mobility~$\mu_\text{TB}$ is thus dominated by the rate~$\Gamma_{1}^{(2)}$
at low temperatures, and vanishes at $T=0$ for $\alpha<1$, which corresponds to free dynamics $\mu=\mu_0$ in the
dual weak-binding system~\cite{note0}. This suggests that the occurrence of a delocalization to localization
transition at~$\alpha=1$ for the ground state of a cosine potential~\cite{FisPRB85,SchPRL83} would not be
affected in more general potentials (see also Fig.~\ref{Fig3}).

In the remainder of the paper, we focus on the ratchet current induced by adiabatic bistable driving
$v_\text{R}^\infty=v^{\infty}_\text{DC}(F)+v^{\infty}_\text{DC}(-F)$. The second-order rates obey
$\Gamma_{m}^{(2)}(-F)=\Gamma_{-m}^{(2)}(F)$ and therefore cancel out in the expression for the ratchet current.
Hence, we have to focus on contributions of at least third order to the rates~$\Gamma_{m}$. Here we neglect
higher orders. This is known to provide a good approximation in TB systems with large dissipation parameter
$\tilde{\alpha}$ and/or high temperature~\cite{WeiBK99}. For simplicity we also consider a potential consisting
of only two harmonics. There is no problem of principle to include more harmonics~\cite{note1}. We find, with
$m=\pm1,\pm2$,
\begin{equation}\label{EqnGamma3rd}
\Gamma_{m}^{(3)}=\frac{V_1^2V_2}{4\hbar^3\gamma^2}\operatorname{Im}\left[\int\nolimits_{-\infty}^{\infty}d\tau \
G_{|m|}^{(3)}(\tau)\ e^{im(F\tilde{L}/\hbar\gamma)\tau-i\operatorname{sgn}(m)\varphi}\right],
\end{equation}
where we have introduced $\varphi=\varphi_2-2\varphi_1$, and
\begin{eqnarray}
\lefteqn{G_1^{(3)}(\tau)=-\int\nolimits_0^{\infty}d\rho\ e^{-2\tilde{\alpha}Q(\rho)}}\nonumber\\
&&\times\left[e^{-2\tilde{\alpha}Q(\tau+\rho)+\tilde{\alpha}Q(\tau+2\rho)}+e^{-2\tilde{\alpha}Q(\tau-\rho)+\tilde{\alpha}Q(\tau-2\rho)}\right],\nonumber\\
\lefteqn{G_2^{(3)}(\tau)=\int\nolimits_0^{\infty}d\rho\ e^{\tilde{\alpha}Q(\rho)-2\tilde{\alpha}Q(\tau+\rho/2)-2\tilde{\alpha}Q(\tau-\rho/2)}.}
\end{eqnarray}
At third order the rates obey
$\Gamma_{m}^{(3)}(F,\varphi)=\Gamma_{-m}^{(3)}(-F,-\varphi)$, which is a consequence of parity. The dependence
of the ratchet current on the potential parameters is then up to third order in the potential amplitude
\begin{equation}
v^\infty_\text{R}\propto V_1^2V_2\sin(\varphi_2-2\varphi_1).
\end{equation}
The ratchet current vanishes for a symmetric potential $\sin{(\varphi_{2}-2\varphi_{1})}=0$ as it should.

\begin{figure}
\includegraphics[angle=270]{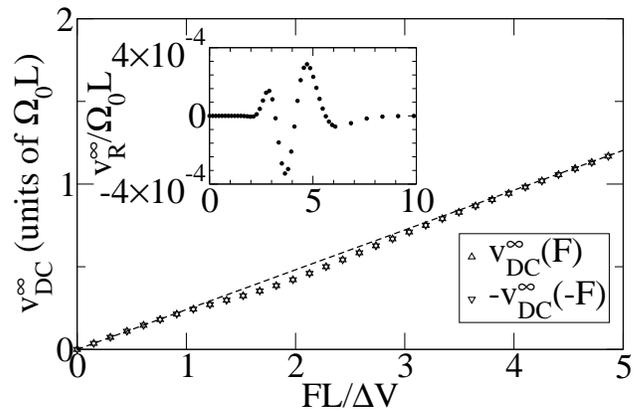} \caption{\label{Fig4}Stationary velocity and ratchet current (inset) as a function of
driving for the potential of amplitude $\Delta V$ depicted in Fig.~\ref{Fig1}. The dashed line is the classical
solution in the absence of potential. Weak dissipation is chosen with $\alpha=0.2$ and $\hbar\gamma=0.76\Delta
V$. Temperature is set to $k_\text{B}T=0.076\Delta V$.}
\end{figure}

The behavior of the particle and ratchet currents as function of temperature and driving is shown in
Figs.~\ref{Fig3} and~\ref{Fig4}
for a two-harmonics potential. In Fig.~\ref{Fig3}, the driving is set to $FL=0.57\Delta V$, whereas in
Fig.~\ref{Fig4}, the temperature is fixed to $k_\text{B}T=0.23\Delta V$. With $V_1=4V_2$, the untilted
potential, depicted in Fig.~\ref{Fig1}, has a barrier height $\Delta V=2.2V_1$. We choose $\alpha=0.2$ and
$\hbar\gamma=0.76\Delta V$. It means that the typical action is $\sqrt{2M\Delta V L^2}\approx2\hbar$, and the
dissipation rate~$\gamma=\eta/M$ is about one-fourth of the classical oscillation frequency
$\Omega_0=2\pi\sqrt{V_1/ML^2}$ in the untilted potential (weak dissipation). In this numerical application, none
of the rates exceeds~$0.05\gamma$ and ~$0.08\omega_{\text{B}}$, which means that the duality relation is valid
for this system. Moreover, the third-order rates stay at least one order of magnitude below the second-order
ones. The ratchet current presents several reversals as a function both of the driving and the temperature. As
expected for the small values of driving and dissipation used in Fig.~\ref{Fig3}, the stationary velocity is
very close to the value of a free system~$v_0=F/\eta$ at $T=0$, which corresponds to \textit{localization}
$v_\text{TB}^{\infty}\approx0$ in the TB system~\cite{note2}. Accordingly, $v_\text{R}^{\infty}\approx0$ in this
regime. The stationary velocity also tends to~$v_0$ (dashed line in Fig.~\ref{Fig4}) for driving or temperatures
much higher than the potential barrier, and the ratchet current vanishes correspondingly. If observed in
experiments, this linear behavior would provide a direct estimation of dissipation.

In conclusion, we obtained a duality relation yielding a tight-binding description of Brownian motion in a
tilted ratchet potential. We demonstrated its application to investigate rectification of adiabatic driving and
ground-state delocalization for weak dissipation.

%
% Acknowledgments
%
\begin{acknowledgments}
We thank U.~Weiss for seminal discussions. This work was supported
by the Dutch Foundation FOM.
\end{acknowledgments}


\begin{thebibliography}{10}
\bibitem{RatRev}Appl. Phys. A {\bf 75}, 167 (2002),
special issue on \textit{Ratchets and Brownian motors}; P.~Reimann, Phys. Rep. {\bf 361}, 57 (2002);
R.~D.~Astumian and P.~H\"anggi, Phys. Today {\bf 55} (11), 33 (2002).
\bibitem{WeiBK99}U.~Weiss, {\em Quantum Dissipative Systems}, 2nd ed.
(World Scientific, Singapore, 1999).
\bibitem{ReiPRL97}P.~Reimann, M.~Grifoni, and P.~H\"anggi,
Phys. Rev. Lett. {\bf 79}, 10 (1997).
\bibitem{LinSci99}H.~Linke {\em et al.}, Science {\bf 286}, 2314 (1999).
\bibitem{MajPRL03}J.~B.~Majer {\em et al.},
Phys. Rev. Lett. {\bf 90}, 056802 (2003).
\bibitem{RonPRL98}R.~Roncaglia and G.~P.~Tsironis,
Phys. Rev. Lett. {\bf 81}, 10 (1998).
\bibitem{SchPRB02}S.~Scheidl and V.~M.~Vinokur,
Phys. Rev. B {\bf 65}, 195305 (2002).
\bibitem{LehPRL02}J.~Lehmann {\em et al.},
Phys. Rev. Lett. {\bf 88}, 228305 (2002).
\bibitem{GriPRL02}M.~Grifoni {\em et al.},
Phys. Rev. Lett. {\bf 89}, 146801 (2002).
\bibitem{MacPRE04}L.~Machura {\em et al.},
Phys. Rev. E {\bf 70}, 031107 (2004).
\bibitem[Fisher and Zwerger(1985)Fisher and Zwerger]{FisPRB85}
M.~P.~A.~Fisher and W.~Zwerger, Phys. Rev. B {\bf 32}, 6190 (1985).
\bibitem{GriPRE96}M.~Grifoni, M.~Sassetti, and U.~Weiss,
Phys. Rev. E {\bf 53}, R2033 (1996).
\bibitem{SasPRB96}M.~Sassetti,
H.~Schomerus, and U.~Weiss, Phys. Rev. B {\bf 53}, R2914 (1996).
\bibitem{note0}Due to the behavior of the rates, the rare transitions limit can hold down to $T=0$ for $\alpha<1$.
\bibitem{SchPRL83}A.~Schmid,
Phys. Rev. Lett. {\bf 51}, 1506 (1983).
\bibitem{note1}A potential with
few harmonics also comes naturally in experiments with arrays of Josephson junctions~\cite{MajPRL03}.
\bibitem{note2}We obtain the opposite behavior
$v_\text{TB}\to\infty$ at low temperatures for $\alpha=1.26$.
\end{thebibliography}
\end{document}